\documentclass[aps,twocolumn,prl,amssymb,citesort]{revtex4}
\usepackage{amsmath}
\usepackage{graphicx}
\usepackage{color}

\begin{document}
\title{\Large Finding unprecedentedly low-thermal-conductivity \\half-Heusler semiconductors via high-throughput materials modeling}

\author{Jes\'us Carrete} \author{Wu Li} \author{Natalio Mingo}
\email{natalio.mingo@cea.fr}
\affiliation{CEA-Grenoble, 17 Rue des Martyrs, Grenoble 38054, France}

\author{Shidong Wang} \author{Stefano Curtarolo}
\email{stefano@duke.edu}
\affiliation{Center for Materials Genomics, Materials Science, Electrical Engineering, Physics and Chemistry, Duke University, Durham, North Carolina 27708, USA}

\begin{abstract}
The lattice thermal conductivity ($\kappa_{\mathrm{\omega}}$) is a key property for many potential applications of compounds. 
Discovery of materials with very low or high $\kappa_{\mathrm{\omega}}$ remains an experimental challenge due to high costs and time-consuming synthesis procedures. 
High-throughput computational pre-screening is a valuable approach for significantly reducing the set of candidate compounds. 
In this article, we introduce efficient methods for reliably estimating the bulk $\kappa_{\mathrm{\omega}}$ for a large number of compounds. 
The algorithms are based on a combination of machine-learning algorithms, physical insights, and automatic {\itshape ab-initio} calculations.
 We scanned approximately {$79,000$} half-Heusler entries in the {\small AFLOWLIB}.org database. 
 Among the $450$ mechanically stable ordered semiconductors identified, we find that $\kappa_{\mathrm{\omega}}$ spans more than two orders of magnitude- a much larger range than that previously thought. 
 $\kappa_{\mathrm{\omega}}$ is lowest for compounds whose elements in equivalent positions have large atomic radii. 
We then perform a thorough screening of thermodynamical stability that allows to reduce the list to $77$ systems. 
We can then provide a quantitative estimate of $\kappa_{\mathrm{\omega}}$ for this selected range of systems. 
{Three} semiconductors having $\kappa_{\mathrm{\omega}} < 5\,\mathrm{W\,m^{-1}\,K^{-1}}$ are proposed for further experimental study.
\end{abstract}
\maketitle

\section{Introduction}

High-throughput (HT) computational materials science is a rapidly expanding area of materials research. 
It merges a plethora of techniques from a variety of disciplines. 
These include the kinetics and thermodynamics of materials, solid-state physics, artificial intelligence, computer science, and statistics \cite{curtarolo:nmat_review}. 
The application of HT has recently led to new insights and novel compounds in different 
fields \cite{Greeley2006,CederMRSB2011,aflowTHERMO,CastelliJacobsen2012_EnEnvSci,YuZunger2012_PRL,curtarolo:TIs,Ceder_ScientificAmerican_2013,monsterPGM}. 
Despite the importance of thermal transport properties for many crucial technologies, there are to date no high-throughput investigations into lattice thermal conductivity.

\begin{figure*}[t!]
 \begin{center}\noindent\includegraphics[width=0.90\linewidth]{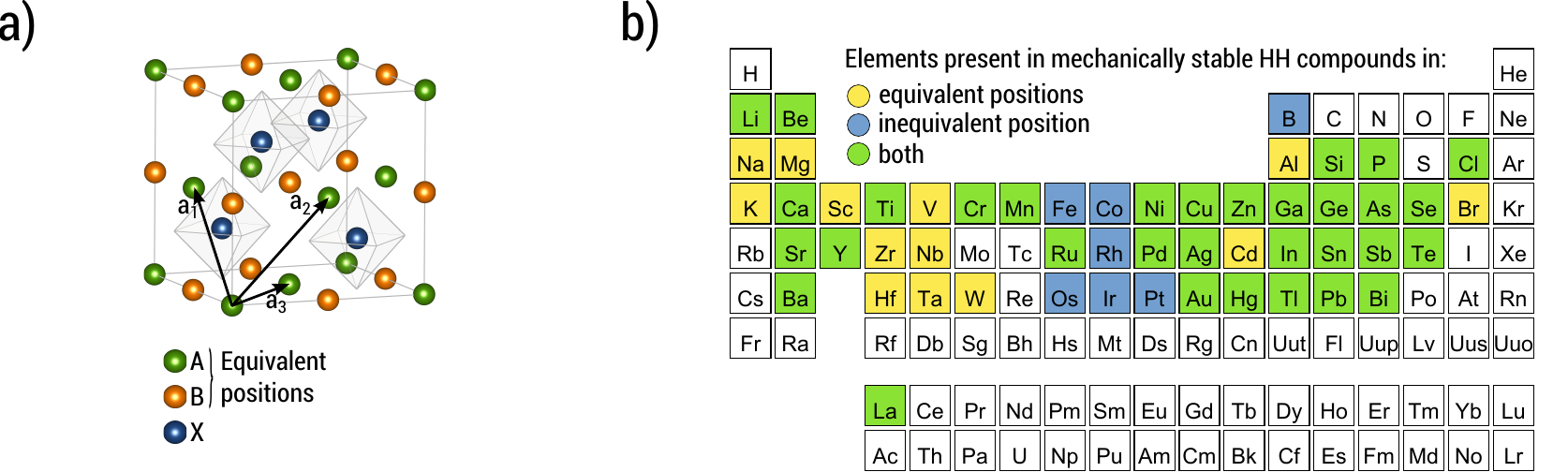}\end{center}
 \caption{\small
  {\bfseries (a)} Prototype Half-Heusler structure with primitive vectors and conventional cell.
  {\bfseries (b)} Elements considered in this study.
}
 \label{fig1}
\end{figure*}

Here we seek to address this challenge. 
We concentrate on the lattice thermal conductivity of half-Heusler compounds, as they have great promise for applications as thermoelectric materials \cite{Uher_PRB_1999,Hohl_JPCM_1999,Nolas_MRS_2006,Yu_ACTAMAT_2009}. 
Half-Heusler compounds are ternary solids. Their crystalline structure consists of two atoms ($A$ and $B$), located in equivalent positions in a rock-salt structure. 
A third atom ($X$) sits in an inequivalent position, filling half of the octahedrally coordinated sites (Fig.~\ref{fig1}a).

Experimental studies have reported the thermoelectric figure of merit for a small set of these systems and their alloys \cite{sekimoto,Culp_APL_2006,Joshi_AEM_2011,Yan_NL_2011,yan_stronger_2012}. 
Theoretical electronic characterizations have been performed for 36 candidates \cite{shiomi_thermal_2011}. 
It has been speculated that their high thermal conductivity, close to $10\,\mathrm{W\,m^{-1}\,K^{-1}}$, could limit thermoelectric performance \cite{casper_half-heusler_2012,xie_recent_2012}. 
At room temperature, the lattice thermal conductivity $\kappa_{\mathrm{\omega}}$ represents the largest contribution to the total conductivity. 

Promising thermoelectric figures of merit have been reported both for $n$-type ($1.5$ at $700\,\mathrm{K}$ \cite{sakurada_effect_2005}) and for $p$-type ($0.8$ at $1000\,\mathrm{K}$ \cite{Yan_NL_2011}) half-Heuslers.
Such values are comparable to the best thermoelectric materials proposed thus far \cite{li_high-performance_2010}. 
Those values, however, were not found in ordered half-Heuslers; but rather in alloyed or nanostructured systems. 
Furthermore, finding ordered compounds with very low $\kappa_{\mathrm{\omega}}$ is advantageous, as their electronic mobilities are expected to be higher than in alloys. 
In addition, alloying the already low-$\kappa_{\mathrm{\omega}}$ ordered compounds would lower $\kappa_{\mathrm{\omega}}$ even further.

The pool of candidate compounds analyzed in this article is larger than in previous investigations. 
All possible half-Heusler compounds from all combinations of non-radioactive elements in the periodic table are considered, as included in the {\small AFLOWLIB}.org consortium repository \cite{aflowlibPAPER} (Fig.~\ref{fig1}b). 
The formation enthalpies of the fully relaxed structures are obtained through density functional theory within the {\small AFLOW} high-throughput framework \cite{aflowPAPER}. 

From a total of $79,057$ entries, those with positive formation enthalpies are removed. 
When several half-Heuslers are related by permutations of elements, only the lowest-enthalpy configurations are considered. 
Finally, zero-gap compounds are removed from the list. 
For the surviving subset of $995$ compounds, the second-order force constants are characterized with full phonon dispersion curves.
This allows further reduction of the set to a total of $450$ mechanically stable semiconductors. 
Although these requirements do not guarantee global thermodynamical stability, metastable compounds with long lifetimes have been synthesized and used \cite{Zhang_ADFM_2012}. 
Hence, their inclusion should not be disregarded {\itshape a priori}.

For the $450$ resulting stable half-Heuslers, we compute a large set of structural, electronic and harmonic properties. 
In principle, one could directly compute $\kappa_{\mathrm{\omega}}$ for all the compounds. 
The computational requirements for this approach would be prohibitive. 
To solve this issue, our strategy is to obtain $\kappa_{\mathrm{\omega}}$ for a smaller subset of systems.
We use physical insights and machine learning techniques to predict the remaining values.
Cross-validation shows that the approach is reliable for rapidly identifying low-$\kappa_{\mathrm{\omega}}$ compounds.

Once the main factors correlated with a low thermal conductivity are identified for the $450$-HH library, we use the thermodynamical information in the {\small AFLOWLIB}.org database to test the stability of these HHs against more than $110,000$ phases. 
All competing ternary compounds from the Inorganic Crystal Structure Database (ICSD) \cite{ICSD_database} and all binaries in that database sharing two elements with each HH are included.
The final list of thermodynamically stable compounds contains $77$ entries. 
For these we devise and implement a novel approach to compute the lattice thermal conductivity. 
Our accuracy is better than $50\%$ of the exact calculation, and has a much lower computational cost. 
This allows to provide estimates of $\kappa_{\mathrm{\omega}}$ that can be compared with experiment for $77$ thermodynamically stable compounds.

\section{Predicting bulk lattice thermal conductivities}
The general expression for $\kappa_{\mathrm{\omega}}$ at temperature $T$ is \cite{ziman_electrons_2001}:
\begin{align}
 \kappa_{\mathrm{\omega}}=\frac{1}{k_BT^2V}\sum\limits_{\lambda}n_0\left(n_0+1\right)\lvert v_{\lambda}^{\left(z \right)}\rvert^2\hbar^2\omega_{\lambda}^2\tau_{\lambda},
 \label{eqkappa}
\end{align}
\noindent where
$\lambda$ denotes the phonon branch index $\alpha$ and wave vector $\mathbf{q}$,
$k_B$ is Boltzmann's constant,
$n_0$ the Bose-Einstein distribution,
$\omega_{\lambda}$ the frequency of phonons,
$v_{\lambda}^{\left(z \right)}$ the phonon group velocity in the transport direction $z$, and
$\tau_{\lambda}$ the relaxation time. The relaxation time is determined by third-order derivatives of the total energy with respect to the atomic displacements of any three atoms $i$, $j$, and $k$ in directions $a$ $b$ and $c$
($\Phi_{ijk}^{abc}$, the anharmonic force constants) in a large supercell \cite{ward_ab_2009}.

In ordered half-Heuslers the dominant source of scattering is due to three-phonon processes, and we can calculate thermal conductivities with the full {\itshape ab-initio} anharmonic characterization \cite{Broido2007,Wu_PRB_2012}.
For CoSbZr, one of the most thoroughly studied {half-Heuslers}, we obtain $\kappa_{\mathrm{\omega}}=25.0\,\mathrm{W\,m^{-1}\,K^{-1}}$.
The value is very close to a previous theoretical estimate ($\sim 22\,\mathrm{W\,m^{-1}\,K^{-1}}$) for monocrystalline CoSbZr \cite{shiomi_thermal_2011}, and
slightly higher than the experimental values \cite{xia_thermoelectric_2000,sekimoto_high-thermoelectric_2007} (between $15$ and $20\,\mathrm{W\,m^{-1}\,K^{-1}}$). 
Synthesized samples of Refs. \cite{xia_thermoelectric_2000,sekimoto_high-thermoelectric_2007} may contain microstructures and imperfections not considered in our work.
Despite of accuracy, ``full  {\itshape ab-initio}  calculations'' of $\kappa_{\mathrm{\omega}}$ (Eq. \ref{eqkappa}) are prohibitive for HT studies, due to the computational requirements of the derivatives giving $\tau_{\lambda}$.

In this section, we present two different approaches circumventing the limitation. 
The first method is based on the empirical observation that the force constants show high degree of transferability between compounds sharing crystal structure \cite{Giannozzi_PRB_1991}.  
This suggests that a single set of anharmonic force constants could be used to get an estimate of the bulk $\kappa_{\mathrm{\omega}}$. We call this thermal conductivity calculated with ``transferred'' forces $\kappa_{\mathrm{transf}}$ {(see Table~\ref{table_notation})}.

\begin{figure}[t!]
  \begin{center}\noindent\includegraphics[width=0.90\linewidth]{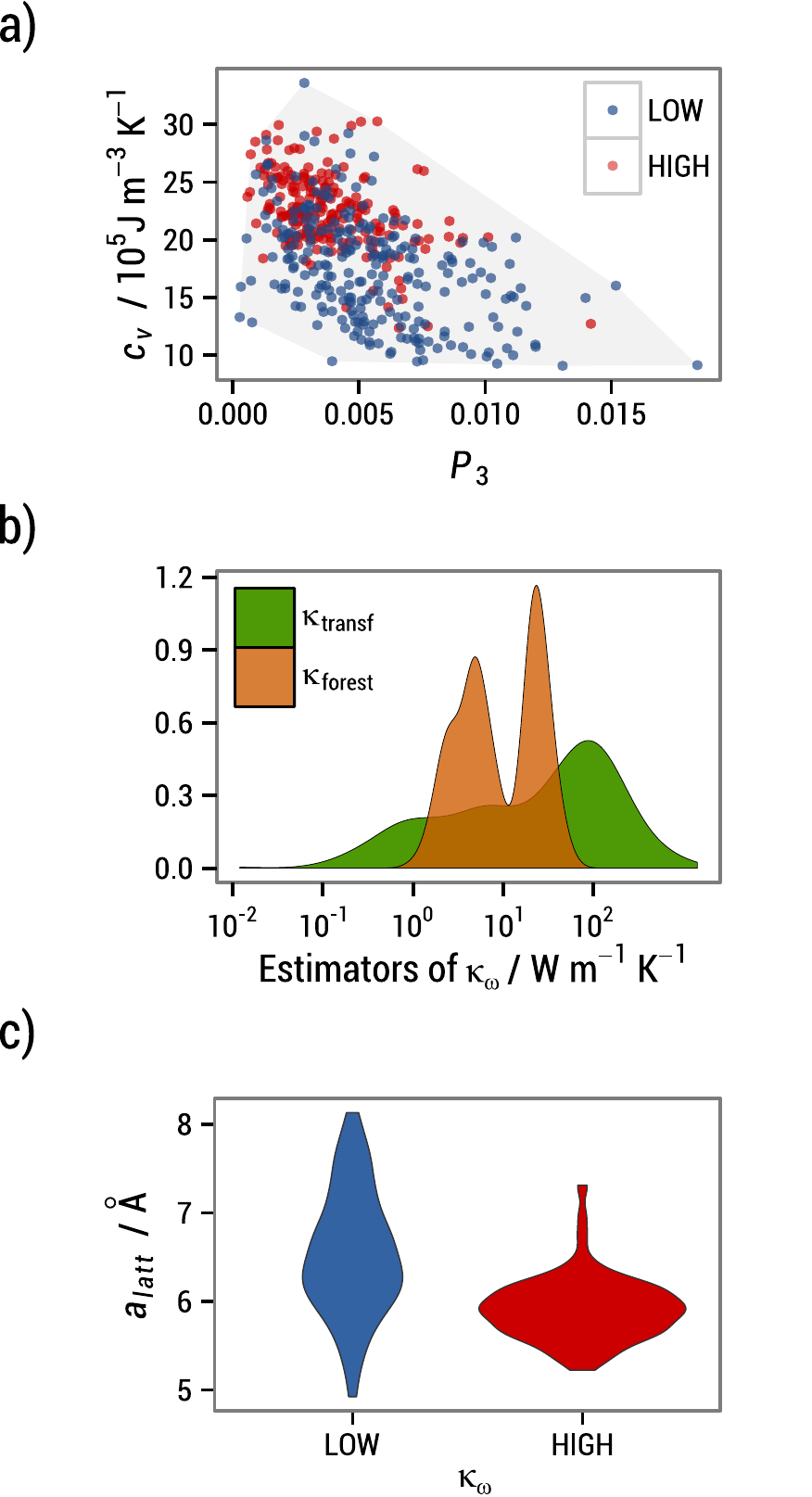}\end{center}
  \vspace{-6mm}
  \caption{\small
    {\bfseries (a)} 
    Joint scatter plot of $c_v$ at $300\,\mathrm{K}$ and $P_3$ colored according to our low/high-$\kappa_{\mathrm{\omega}}$ 
    classification based on $\kappa_{\mathrm{forest}}$ (see text); the convex hull of the point set is also included for guidance.
    {\bfseries {(b)}}
    Frequency densities of the estimators of thermal conductivity at $300\,\mathrm{K}$ $\kappa_{\mathrm{transf}}$ and $\kappa_{\mathrm{forest}}$ 
    as defined in the text.
    {\bfseries (c)} {``Violin plot'' showing the distribution} of $a_{\mathrm{latt}}$ within the {low- and high}-$\kappa_{\mathrm{\omega}}$ classes.
  }
  \label{fig2}
\end{figure}

We desired to preserve the choice between equivalent positions for maximizing transferability. 
Thus, instead of taking the anharmonic force constants of a particular half-Heusler compound, we choose those of Mg$_2$Si.
This compound shares the half-Heusler lattice with sites $A$ and $B$ occupied by Mg atoms.
For cross-validation, we also fully compute the anharmonic force constants of $32$ half-Heusler systems. 
These are randomly selected with uniform probability inside the convex hull of Fig.~\ref{fig2}a, to ensure a wide variety of harmonic/anharmonic features.
Comparison between $\kappa_{\mathrm{\omega}}$ and $\kappa_{\mathrm{transf}}$ indicates that- although the latter has limited quantitative precision- the qualitative agreement is very good; 
with a Spearman rank correlation coefficient of $0.93$. 
Hence, the descriptor can be effectively used to separate compounds having high or low $\kappa_{\mathrm{\omega}}$.  
Note that we chose the Spearman rank correlation \cite{spearman_rank} instead of the usual Pearson's.
The former is invariant under any monotonic transformation of one or both variables, and takes values $\pm 1$ for any strict monotonic (not just linear) dependence.

\begin{table*}[t!]
  \begin{center}\noindent\includegraphics[width=0.90\linewidth]{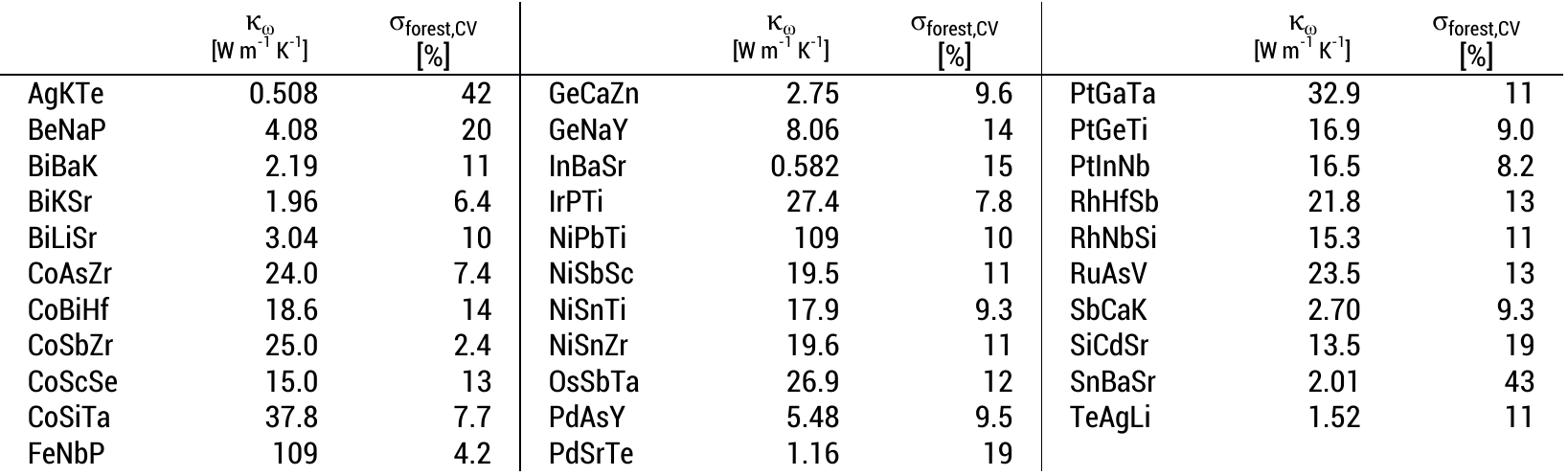}\end{center}
  \vspace{-5mm}
  \caption{\small Fully calculated thermal conductivities, $\kappa_{\mathrm{\omega}}$, for $32$ compounds. 
    These results are then used as the training set for the random-forest predictions. 
    An estimate of the relative standard deviation of $\kappa_{\mathrm{forest}}$ for each compound in the training set, as obtained using repeated $4$-fold cross-validation, is also included.
    Compounds are always labeled with the element in position $X$ first.}
  \label{table1}
\end{table*}

\begin{table}[htb!]
  \begin{center}\noindent\includegraphics[width=0.90\linewidth]{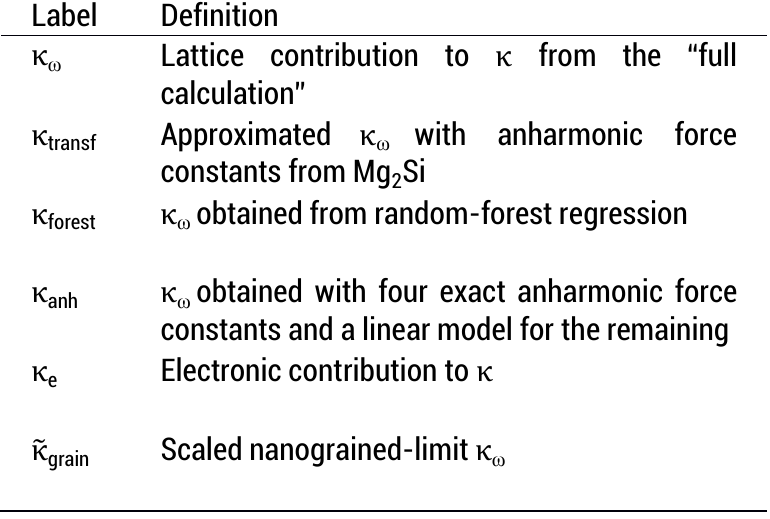}\end{center}
  % \begin{tabular}{l|l}
  %   {\it label}   & {\it definition} \\ \hline
  %   $\kappa_{\mathrm{\omega}}$ & lattice contribution to $\kappa$ from the ``full calculation''  \\
  %   $\kappa_{\mathrm{transf}}$ & approximated $\kappa$ with anharmonic force constants from Mg$_2$Si  \\
  %   $\kappa_{\mathrm{forest}}$ & $\kappa$ obtained from random-forest regression  \\
  %   $\kappa_{\mathrm{anh}}$ & $\kappa$ obtained with four exact anharmonic force constants and a linear model for the remaining   \\
  %   $\kappa_e$ & electronic contribution to $\kappa$  \\
  %   $\tilde{\kappa}_{\mathrm{grain}}$ & scaled nanograined-limit $\kappa$  \\ 
  % \end{tabular}
  \vspace{-5mm}
  \caption{\small
    Notation for thermal conductivies.} 
  \label{table_notation}
\end{table}

The second proposed approach is based on a completely different direction: we use ``random-forest regression'' by leveraging the {$32$} fully calculated $\kappa_{\mathrm{\omega}}$ as a training set.
We can then employ the fitted model to predict the remaining conductivities. We call these predictions $\kappa_{\mathrm{forest}}$ (see Table \ref{table_notation}).
Random forests \cite{randomforests} are a general classification and regression algorithms and the are well adapted to dependent input data. 
They have already been successfully applied to numerous problems \cite{palmer_random_2007,auret_unsupervised_2010}, including compound classification \cite{svetnik_random_2003}. 
Here, the $32$ compounds represent only around $7\%$ of the mechanically stable half-Heuslers.
Our input data {comprises} a large set of descriptor variables, which is expected to correlate with $\kappa_{\mathrm{\omega}}$ (supplementary materials) but is less expensive to obtain.
Descriptors include:

\begin{itemize}
\item
  \textit{A priori} chemical information: 
  atomic number and weight, position in the periodic table, atomic radius,
  Pauling electronegativity \cite{pauling_bond}, and Pettifor's chemical scale $\chi$ (Ref. \onlinecite{pettifor:1984}).
\item
  General compound information: lattice constant {$a_{\mathrm{latt}}$}, band gap, formation {enthalpy}, 
  effective masses of electrons and holes, Born effective charges and dielectric tensor.
\item
  Specific thermal conductivity information: specific heat $c_v$, spherically-averaged speed of sound $c_s$, 
  scaled nanograined-limit thermal conductivity $\tilde{\kappa}_{\mathrm{grain}}$ and phase-space volume available for three-phonon scattering processes $P_3$.
\end{itemize}

\noindent After an exploratory phase, we conclude that a satisfactory fit can be safely achieved using only \textit{a priori} data.

The random-forest method is performed in three steps.
First, a large ensemble of decision trees is built by randomly selecting subsets of descriptors and observations.
Second, the predictions of all trees are obtained for each data point.
Third, the mode (for classification) or the mean (for regression) are taken as the result from the whole ensemble.
The algorithm also provides an intrinsic metric to evaluate the importance of each descriptor.
This is defined in relation to the effect of randomly permuting the values of that variable on the result \cite{randomforests} (the less resilient upon permutation, the more important).

The prediction of each tree in a random forest can only be a value from the training set, and thus the result of the regression is a weighted average.
This average is bounded by the minimum and maximum values within the training data.
A small set is unlikely to contain elements having extreme values.
Hence, our random-forest regression is expected to have a marked centralizing effect, yielding values tightly grouped around their mean.
The frequency densities of both $\kappa_{\mathrm{transf}}$ and this new $\kappa_{\mathrm{forest}}$ are displayed in Fig.~\ref{fig2}b.
The latter avoids extreme predictions with non-physical magnitudes, a result of the aforementioned centralizing effect. 

In this sense, machine-learning algorithms outperform crude extrapolations such as those behind $\kappa_{\mathrm{transf}}$.
Additionally, $\kappa_{\mathrm{forest}}$ has the advantage that its predictions can be refined with controlled accuracy by changing the size of the training set.
Even so, the Spearman rank correlation coefficient between $\kappa_{\mathrm{transf}}$ and $\kappa_{\mathrm{forest}}$ is still $0.66$, corroborating the validity of the analysis based on $\kappa_{\mathrm{forest}}$.
Furthermore, we find that $\kappa_{\mathrm{forest}}$ is strongly correlated with physical descriptors like $c_v$, $\tilde{\kappa}_{\mathrm{grain}}$, and $P_3$.
This confirms our earlier speculation about these methods.

An important concern when training a machine-learning model is whether the training set is diverse or representative enough to justify extrapolating the model to the remaining elements.
The values of $\kappa_{\mathrm{\omega}}$ needed for direct validation of the predicted $\kappa_{\mathrm{forest}}$ are unavailable.
Thus, we resort to a repeated $4$-fold cross-validation among the data points in the training set to obtain an estimate of the out-of-sample error.
More specifically, we evenly split our training set into $4$ subsets.
Then, we obtain a random-forest prediction for the HHs in each of the subsets by using only the remaining $75\%$ of compounds as the new training set.
We repeat the process $10$ times for different divisions of the data, and compute the standard deviation of these predictions.
The results are included in Table.~\ref{table1}.
These estimates support the notion that the model behind $\kappa_{\mathrm{forest}}$ is reasonably insensitive to our choice of training sets. 
{For each cross-validation, we compute the Spearman rank correlation coefficient between the out-of-sample random-forest results for the $32$ training compounds and their $\kappa_{\mathrm{\omega}}$. 
  The median value of these Spearman rank correlation coefficients is $0.74$, corroborating  $\kappa_{\mathrm{forest}}$ as a reliable tool for predicting compounds' ordering.}

The ordering predicted by descriptor {$\kappa_{\mathrm{forest}}$} is strongly correlated with that of $\kappa_{\mathrm{\omega}}$.
This allows to pinpoint the main factors determining high or low thermal conductivities.
The bimodal shape of the distribution in Fig.~\ref{fig2}b suggests that two groups of half-Heuslers can be identified, with thermal conductivities spread around two different values.
A robust version of the ``$k$-means'' algorithm \cite{k_means} is employed to optimally place the medians of the low- and high-thermal-conductivity classes at $4.50$ and $23.1\,\mathrm{W\,m^{-1}\,K^{-1}}$, respectively.
By analyzing the importance of variables in the classification, we identify a low Pettifor scale $\chi_X$ and a large average Pauling electronegativity $\bar{e}_{AB}$ as the most critical descriptors for low conductivity (supplementary materials). 

Given the underlying correlations, many different choices can be used for the classification.
A trend can even be suggested on the grounds of atomic radii by following a chain of correlations: if the two elements in equivalent positions are chosen so that their average radius is larger than {$150\,\mathrm{pm}$}, then the probability of the compound being in the low-$\kappa_{\mathrm{\omega}}$ class is {$84\%$}.
Physically, this follows from the fact that $\kappa_{\mathrm{\omega}}$ is highly correlated with the specific heat $c_v$ (Fig.~\ref{fig2}a).
{The latter} is strongly negatively correlated with the lattice parameter $a_{\mathrm{latt}}$: the larger $a_{\mathrm{latt}}$ the lower $c_v$ 
\footnote{The decreasing trend can be understood considering that the specific heat per atom at high temperatures relates to the number of degrees of freedom,
  through the equipartition theorem. Hence a $c_v\propto a_{\mathrm{latt}}^{−3}$ dependence should be expected.
  In fact, the observed trend is sharper, due to the differences in Debye temperature among the compounds.}. 

In addition, $a_{\mathrm{latt}}$ correlates well with the sum of the atomic radii of the three elements, quantities known {\itshape a priori}.
The atomic radii of the species in positions $X$ concentrate around the average value.
This leads to an accurate prediction of $a_{\mathrm{latt}}$ by using only the average atomic radius of atoms in positions $A$ and $B$, $\bar{r}_{AB}$.
A large $\bar{r}_{AB}$ causes a large lattice constant, small specific heat, and finally, low thermal conductivity.
Alternatively, the lattice parameter can be used as a good discriminant: panel (c) in Fig.~\ref{fig2} is a ``violin plot'' illustrating the distribution of $a_{\mathrm{latt}}$ in the classes of half-Heuslers with low and high thermal conductivities.
Also, as it can be seen in Fig.~\ref{fig2}a our choice of easily computable descriptors such as $c_v$ and $P_3$ is supported by the result of this classification.

Our calculations are for the true bulk lattice thermal conductivity. 
They are unrelated to the minimum value proposed by other authors \cite{slack,cahill_lower_1992}.
Nevertheless, some of the $\kappa_{\mathrm{\omega}}$ obtained directly seem ultra-low.
They are even lower than $\sim 0.70\,\mathrm{W\,m^{-1}\,K^{-1}}$, as reported in literature for AgSbTe$_2$ and AgBiSe$_2$ \cite{morelli_intrinsically_2008}, and described as close to the achievable minimum.
However, the minimum depends on the compound's structure.
Even within the most stringent hypothesis of the shortest possible mean free path equal to interatomic spacing, the lowest found $\kappa_{\mathrm{\omega}}$ is much higher than the theoretical minimum.
Therefore, none of our predicted values violates the minimum lattice thermal conductivity.
Note also that, once the goal of reducing the $\kappa_{\mathrm{\omega}}$ under $\lesssim 1\,\mathrm{W\,m^{-1}\,K^{-1}}$ is achieved, its precise value loses relevancy as it is overtaken by the contribution of charge carriers, $\kappa_e$.

\section{Screening for thermodynamical stability}

The ingredients of $\kappa_{\mathrm{\omega}}$ for bulk ordered semiconductors depend only on a semilocal characterization of the potential energy surface around the equilibrium configuration.
Hence, mechanical stability is sufficient to permit the calculation of the lattice thermal conductivity of a HH.
For the analysis performed in the previous section, having the set of $450$ mechanically stable HHs reduced and biased by external considerations such as thermodynamical
stability would be detrimental to the performance of machine-learning techniques.

On the other hand, in order to propose particular candidates for experimentation we must maximize the probability that they can be obtained in the laboratory.
To this end, we obtain the ternary phase diagrams for each of the $450$ mechanically stable HHs.
This involves taking into account the formation enthalpies of a large number of possible competing phases.
These include but are not limited to all relevant binary and ternary compounds in the ICSD \cite{ICSD_database}.
More specifically, all the elemental compounds, $109,136$ binary structures and $4,363$ ternary phases were considered.
Many of these phases were already present in {\small AFLOWLIB}.org; others were computed specifically for this work.
The total number of DFT calculations necessary to obtain the results presented here exceeds $300,000$.
Our thermodynamic analysis reveals that $79$ of the $450$ HHs are thermodynamically stable. 
{Spin-polarized calculations reveal that two of the $79$ have semimetallic ground states.
Then, only the remaining $77$ compounds are further considered.}
The ternary phase diagrams of the final $77$ systems are included in the supplementary materials.

\begin{figure}[htb!]
  \begin{center}\noindent\includegraphics[width=0.99\linewidth]{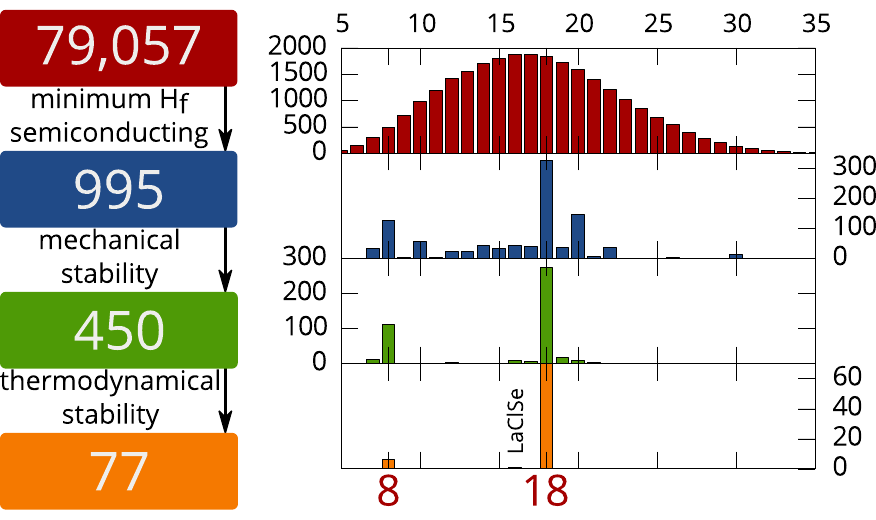}\end{center}
  \vspace{-5mm}
  \caption{\small
    Number of compounds during the screening (left) and evolution of the valence per unit cell distribution (right). {Of the final 77 compounds, only LaClSe (valence=16) does not seem to follow the $8/18$ octet rule.}} 
  \label{fig3}
\end{figure}

$76$ of the $77$ predicted stable compounds satisfy the octet or expanded octet rules by virtue of having $8$ or $18$ valence electrons per unit cell, respectively.
We compared these numbers with the frequency distribution of valence electron counts in the initial $79,057$-HH library.
We conclude that the conditional probabilities of compounds having $8$ or $18$ valence electrons per unit cell being stable are $1.2\%$ and $3.8\%$, respectively.
While still small, the conditional likelihood of a compound satisfying one of these rules making it through 
all the filtering steps is much higher than the $0.1\%$ {\itshape a priori} probability.
Fig. \ref{fig3} shows the distribution of the valence during the reduction of prototypes' list. 
{Only the compound LaClSe (valence=16) does not seem to follow the rule.}

Even among the reduced list, $\kappa_{\mathrm{forest}}$ still spans more than one order of magnitude, 
its extreme values being $2.33$ and $40.3\,\mathrm{W\,m^{-1}\,K^{-1}}$, reinforcing our previous conclusions.

\section{A descriptor with quantitative power}

Neither of the two descriptors of $\kappa_{\mathrm{\omega}}$ presented so far contains any information about the anharmonic interatomic force constants (IFCs) of each compound.
On one hand, the last round of thermodynamical screening puts the number of surviving HHs within the limits of what can be realistically considered for anharmonic calculations.
On the other, the qualitative success of $\kappa_{\mathrm{transf}}$ shows that a detailed anharmonic description is not required.
To enhance our estimates of the thermal conductivity of stable half-Heuslers, in this section we present a new machine-learning descriptor of $\kappa_{\mathrm{\omega}}$ that integrates only the crucial pieces of the anharmonic properties of the solid.
This aids in achieving quantitative accuracy with a much lower computational cost than the full calculation.

Crystallographic symmetries and the equality of mixed partials impose linear constraints on the anharmonic IFCs.
With the parameters described in the ``Methods'' section below and those constraints, we are left with $737$ independent anharmonic IFCs per compound.
However, many elements of this set are correlated among them, and others are too small to have a decisive role in the value of $\kappa_{\mathrm{\omega}}$.
To quantify these assertions, we perform principal component analysis \cite{jolliffe_principal_2002} on the third-order IFCs for the $32$ compounds in Table.~\ref{table1}. 

We find that the first four components account for $\sim 99\%$ of the variance in the set.
From the results we can extract an expression for each of the $737$ IFCs as a linear combination of those components.
Then we perform a multivariable multiple linear regression of the four components on four large and weakly correlated IFCs.
By combining the two results, we arrive at a linear model for the whole set of anharmonic IFCs in terms of four parameters that can be obtained with $16$ DFT 
calculations per compound.
We use the term $\kappa_{\mathrm{\omega}}$ to describe the third-order IFCs thus reconstructed, and $\kappa_{\mathrm{anh}}$ for the second-order IFCs for each compound.

\begin{figure}[htb!]
  \begin{center}\noindent\includegraphics[width=0.90\linewidth]{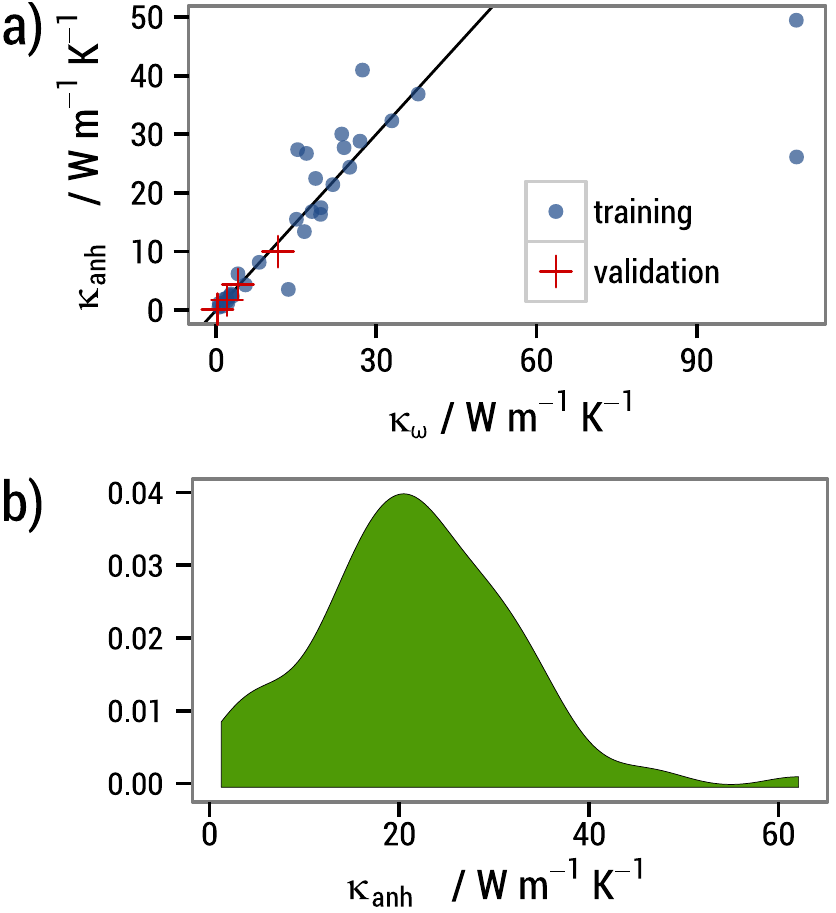}\end{center}
 \vspace{-5mm}
  \caption{\small
    {\bfseries (a)} Comparison of $\kappa_{\mathrm{anh}}$ with the exact $\kappa_{\mathrm{\omega}}$ for the $32$ compounds in the training set and the three compounds used for validation.
    {\bfseries (b)} Distribution of $\kappa_{\mathrm{anh}}$ over the $77$ thermodynamically stable HHs.}
  \label{fig4}
\end{figure}

\begin{table*}[htb!]
 \begin{center}\noindent\includegraphics[width=0.80\linewidth]{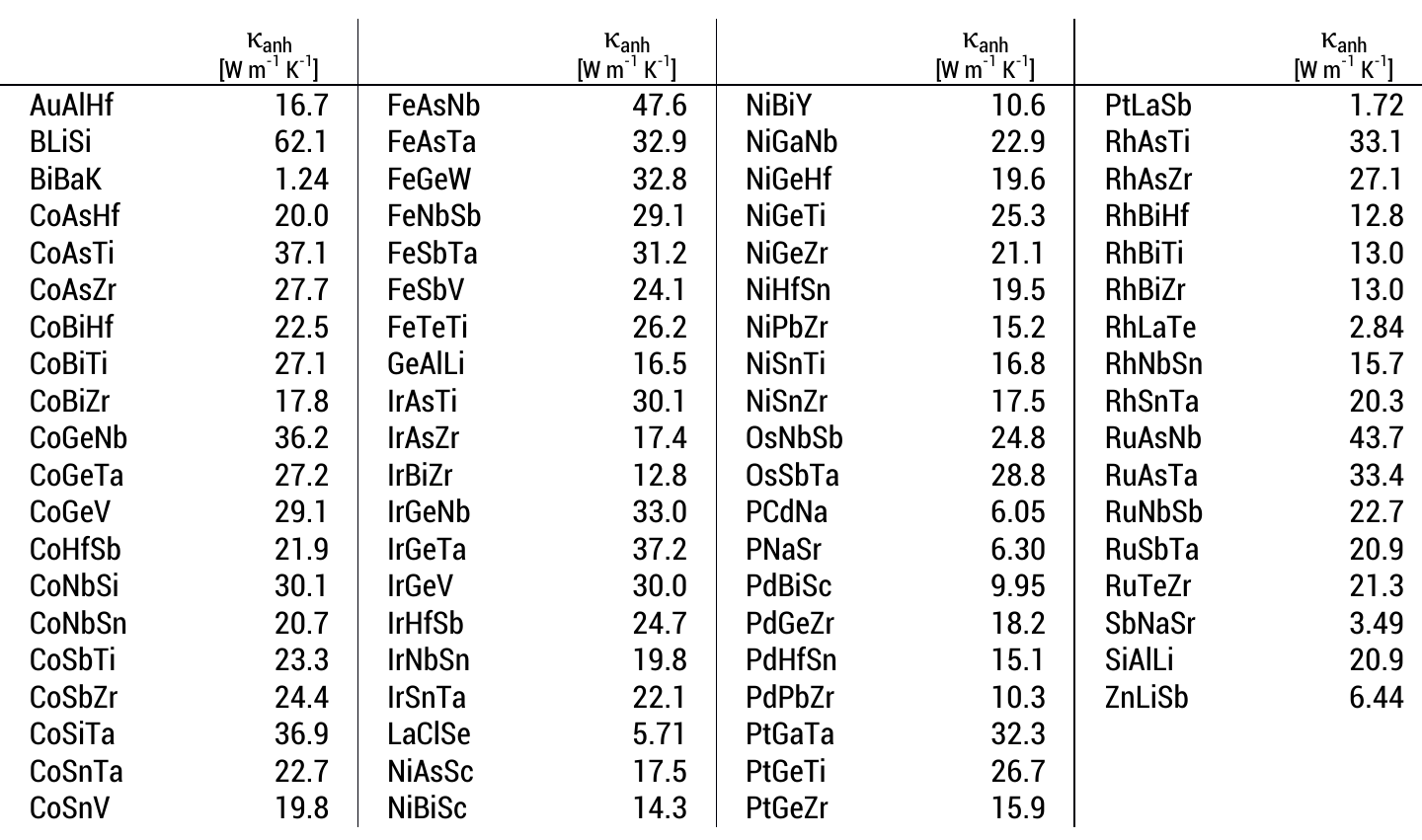}\end{center}
 \vspace{-5mm}
  \caption{\small
    The values of $\kappa_{\mathrm{anh}}$ for the $77$ thermodynamically stable half-Heusler compounds.}
  \label{table3}
\end{table*}

The {blue} circles in Fig.~\ref{fig4}a show a comparison between $\kappa_{\mathrm{anh}}$ and the exact $\kappa_{\mathrm{\omega}}$ for the $32$ compounds in the training set.
With two exceptions (compounds with comparatively very high thermal conductivities) this new descriptor yields excellent quantitative estimates of {$\kappa_{\mathrm{\omega}}$}.
Moreover, $4$-fold cross-validation shows that it is insensitive to the particular choice of training set.
As a final test, we perform full thermal conductivity calculations for four compounds selected at random from those outside the training set:
AgBaSb, AgNaTe, InCdY and TlLaMg.
The results are depicted as {red crosses} in Fig.~\ref{fig4}a.
This shows that the quality of the prediction is as good as for the $32$ training compounds.

The distribution of $\kappa_{\mathrm{anh}}$ over the $77$ thermodynamically stable HHs (Fig.~\ref{fig4}b) confirms the presence in the sample of 
compounds with thermal conductivities much lower that the $10-20\,\mathrm{W\,m^{-1}\,K^{-1}}$.
This is characteristic of experimentally measured HHs. The values of $\kappa_{\mathrm{anh}}$ for the $77$ stable HHs is listed in Table~\ref{table3}.
Notably, the subset of $10$ thermodynamically stable half-Heuslers for which $\kappa_{\mathrm{\omega}}$ was directly computed already contains BiBaK,
with $\kappa_{\mathrm{\omega}}=2.20\,\mathrm{W\,m^{-1}\,K^{-1}}$.
{Outside of the training sample, the lowest $\kappa_{\mathrm{anh}}$ values are $1.72$, $2.84$ and $3.49$ for 
PtLaSb, RhLaTe and SbNaSr, respectively.}

\section{Conclusions}

In this article, we have presented three computational methods for estimating the bulk $\kappa_{\mathrm{\omega}}$ of a large library of half-Heusler compounds.
We surmount the formidable task of full {\itshape ab-initio} characterization.
We find that $\kappa_{\mathrm{\omega}}$ is spread over more than two orders of magnitude over mechanically stable half-Heuslers.
This is a much broader range than that suggested by limited experimental available data.
By using a set of descriptors and random-forest regression, we have built and tested an effective classification model.
We found that compounds are most likely to have low thermal conductivity if the average atomic radius of the atoms in structural positions $A$ and $B$ is large.
This also correlates with large lattice parameters and low specific heats.

Extensive thermodynamical calculations allow to remove from the list compounds with more stable competing phases.
We employ our third method, with better quantitative accuracy and higher computational cost, to perform a finer analysis of the distribution of $\kappa_{\mathrm{\omega}}$ over the reduced library.
We conclude ordered half-Heusler compounds with $\kappa_{\mathrm{\omega}}\lesssim 3\,\mathrm{W\,m^{-1}\,K^{-1}}$ value (a factor of three below the best scenarios for ordered compounds, and comparable to alloyed systems) very likely exist.
The results corroborate the competitiveness of machine-learning methods in accelerated material design \cite{curtarolo:nmat_review}.

\section{Methods}
{\bfseries AFLOWLIB library of half-Heusler systems.}
The $79,057$ half-Heusler systems are calculated with the high-throughput framework {\small AFLOW} \cite{aflowTHERMO,aflowPAPER,aflowBZ,aflowSCINT} based on {\itshape ab-initio} calculations of the energies by the {\small VASP} software \cite{VASP} with projector augmented waves (PAW) pseudopotentials \cite{PAW}, and Perdew, Burke and Ernzerhof exchange-correlation functionals \cite{PBE}.
 The {\small AFLOWLIB} energies are calculated at zero temperature and pressure, with spin polarization and without zero-point motion or lattice vibrations.
All crystal structures are fully relaxed (cell volume and shape and the basis atom coordinates inside the cell). 
Numerical convergence to about $1\,\mathrm{meV\,atom^{-1}}$ is ensured by a high energy cutoff (30\% higher than the highest energy cutoff for the pseudo-potentials of the components) and 
by the dense 6,000 {\bfseries k}-points per reciprocal atom Monkhorst-Pack meshes \cite{monkhorst}.

{\bfseries Interatomic force constants.}
$3\!\times\!3\!\times\!3$ supercells are used in second-order IFC calculations. The Phonopy \cite{phonopy} package is used to generate a minimal set of atomic displacements by harnessing the point and translational symmetries of the crystal structure, and custom software was developed in order to do the same in anharmonic IFC calculations. For those, $4\!\times\!4\!\times\!4$ supercells are generated and a cutoff radius of $0.85a_{\mathrm{latt}}$ is imposed on the interactions. $2\!\times\!2\!\times\!2$ and $3\!\times\!3\!\times\!3$ Monkhorst-Pack {\bfseries k}-point grids are employed, and spin polarization excluded to improve speed.

{\bfseries Solution of the {Boltzmann transport equation}.}
Our self-consistent iterative approach is described in detail in Ref. \onlinecite{Broido2007}. Both three-phonon processes and the natural isotopic distribution of each element are taken into account as source{s} of scattering. A Gaussian smearing scheme with adaptive breadth \cite{Wu_PRB_2012} is chosen for integrations in the Brillouin zone. When using anharmonic IFCs from Mg$_2$Si to approximate $\kappa_{\mathrm{\omega}}$ for all materials, the solution to the {Boltzmann transport equation} failed to converge for $5$ compounds, which are consequently excluded from the associated analysis.

{\bfseries Regression and classification.}
The R statistical computing environment \cite{R} is chosen for all statistical analyses. Random-forest models are used as implemented in the ``randomForest'' package \cite{randomForest}.
As a check, all regressions and classifications are repeated using a generalized boosted tree algorithm \cite{gbm}; in all cases the results are found to be in good agreement with those afforded by random forests.

\section{Acknowledgments}

\begin{acknowledgments}
  The authors thank Prof. D. Broido and Dr. L. Lindsay for providing us with a set of anharmonic force constants for Mg$_2$Si, and Dr. A. Stelling, Dr. O. Levy, Prof. S. Sanvito, Prof. M. Buongiorno Nardelli, and Prof. M. Fornari for useful comments. 
  This work is partially supported by the {French ``Carnot'' project SIEVE,} by DOD-ONR (N00014-13-1-0635, N00014-11-1-0136, N00014-09-1-0921) and by the Duke University---Center for Materials Genomics.
\end{acknowledgments}

\end{document}